# Validating remotely sensed biomass estimates with forest inventory data in the western US

Xiuyu Cao [a], Joseph O. Sexton [b], Panshi Wang [b], Dimitrios Gounaridis [a], Neil H. Carter [a], Kai Zhu [a]

[a] School for Environment and Sustainability, University of Michigan, Ann Arbor, MI, 48109, USA

[b] terraPulse, Inc. 13201 Squires Court, Gaithersburg, MD, 20878, USA

Corresponding author: Kai Zhu (zhukai@umich.edu)

# Abstract

Monitoring aboveground biomass (AGB) and its density (AGBD) at high resolution is essential for carbon accounting and ecosystem management. While NASA's spaceborne Global Ecosystem Dynamics Investigation (GEDI) LiDAR mission provides globally distributed reference measurements for AGBD estimation, the majority of commercial remote sensing products based on GEDI remain without rigorous or independent validation. Here, we present an independent regional validation of an AGBD dataset offered by terraPulse, Inc., based on independent reference data from the US Forest Service Forest Inventory and Analysis (FIA) program. Aggregated to 64,000-hectare hexagons and US counties across the US states of Utah, Nevada, and Washington, we found very strong agreement between *terraPulse* and FIA estimates. At the hexagon scale, we report $R^2 = 0.88$, RMSE = 26.68 Mg ha$^{-1}$, and a correlation coefficient (r) of 0.94. At the county scale, agreement improves to $R^2 = 0.90$, RMSE = 32.62 Mg ha$^{-1}$, slope = 1.07, and r = 0.95. Spatial and statistical analyses indicated that *terraPulse* AGBD values tended to exceed FIA estimates in non-forested areas, likely due to FIA's limited sampling of non-forested vegetation. The *terraPulse* AGBD estimates also exhibited lower values in high-biomass forests, likely due to saturation effects in its optical remote-sensing covariates. This study advances operational carbon monitoring by delivering a scalable framework for comprehensive AGBD validation using independent FIA data, as well as a benchmark validation of a new commercial dataset for global biomass monitoring.

Keywords: biomass monitoring, remote sensing, validation, Forest Inventory and Analysis (FIA)

# 1 Introduction

## 1.1 Background

Aboveground biomass (AGB, in Mg) and its density (AGBD, Mg ha$^{-1}$) represent the total mass of living vegetation above the soil surface and serve as key indicators of carbon stocks and vegetation dynamics (Ashton et al., 2012; Baccini et al., 2012; Davidson & Janssens, 2006; Vashum & Jayakumar, 2012). High-resolution and spatio-temporally explicit estimation of AGB and AGBD is essential for precise carbon accounting and detailed ecosystem management decision making (Friedlingstein et al., 2025; Pan et al., 2011). While *in situ* field measurements of AGBD provide the highest accuracy, satellite remote sensing has become the primary method for large-scale biomass estimation due to its extensive spatial coverage and operational efficiency (Chamberlin et al., 2024; Goetz et al., 2009; Lu et al., 2016).

NASA's Global Ecosystem Dynamics Investigation (GEDI) mission—the first spaceborne LiDAR explicitly designed for vegetation monitoring—has enabled significant advances in AGBD mapping; however, key limitations persist. Although GEDI provides waveform-based structural data at ~25 m footprint resolution, sampled every ~60 m along its orbital track (Duncanson et al., 2022), its coverage is restricted to 51.6°N–S latitudes, and its historical coverage extends only to 2019 (R. Dubayah et al., 2022). Additionally, imputation is required to generate wall-to-wall spatial coverage. Many resulting AGBD products are limited to coarse spatial resolutions (~1 km), single-year snapshots, or site-specific studies (R. Dubayah et al., 2022; Shendryk, 2022; Sialelli et al., 2024; Tamiminia et al., 2024; Zurqani, 2025). Some products enhance spatial resolution by integrating GEDI with other satellite-derived covariates, but they often lack validation against independent ground-based datasets (Sialelli et al., 2024; Zurqani, 2025). This limitation in validation would propagate the bias in the modeled reference data and lead to unreliable assessment (Chave et al., 2019). While some studies have used independent field data for validation (Shendryk, 2022), they still tend to overlook spatial variability in the model's performance, relying instead on metrics such as Pearson's correlation coefficient and similar statistics. This omission can obscure important regional differences in model performance—variation that is ecologically meaningful across heterogeneous landscapes (Lu et al., 2016).

Complementarily, the U.S. Forest Service's Forest Inventory and Analysis (FIA) program offers a unique opportunity for benchmarking AGBD estimates. FIA provides systematically collected, field-based biomass measurements across the United States, with records extending back decades (Burrill et al., 2021). However, its utility for validation is constrained by perturbation of plot locations through "fuzzing" (typically within 0.8–1.5 km) and coordinate "swapping" of up to 20% of plots on private lands within counties (Lister et al., 2005), which preclude point-level comparisons with remote sensing products.

Although prior studies have aggregated FIA estimates to larger spatial units to reduce location error effects, they have encountered several limitations. Firstly, there is a reported potential for temporal mismatch between aggregated FIA AGBD and remote-sensing AGBD due to FIA's small sample sizes within each aggregation unit (Shendryk, 2022). Secondly, although there are available aggregated FIA datasets, they provide only a one-time snapshot of AGBD estimates (R. Dubayah et al., 2022; Menlove & Healey, 2020). This neglects changes in AGBD over time and hinders the validation of future AGBD maps. Therefore, a replicable validation framework that rigorously accommodates spatial uncertainty while enabling independent comparison is thus needed to assess the accuracy of emerging high-resolution AGBD products—especially commercial ones—and to support their use in operational carbon monitoring.

## 1.2 Objectives

This study provides an independent, spatially explicit validation of a commercial AGBD dataset of annual, 30-m resolution estimates of AGBD from 1984 to the present produced by terraPulse, Inc. (https://www.terrapulse.com) against temporally coincident but spatially coarsened reference measurements from the FIA program. We compared *terraPulse* estimates against FIA-based AGBD values, aggregated to 64,000-ha hexagons and U.S. counties across a biomass gradient spanning the US states of Utah, Nevada, and Washington. This validation framework accounts for FIA's location uncertainty while enabling regionally specific evaluation across multiple time points. Our approach delivers both a benchmark validation of a commercial remote sensing product and a replicable framework for assessing future biomass datasets using independent field references.

# 2 Methods

## 2.1 Study area

Our validation was conducted across three western U.S. states—Utah, Nevada, and Washington—selected to span the range of AGBD values from deserts to temperate rainforests (Figure 1). Utah and Nevada are characterized by arid to semi-arid temperate ecosystems, with mean annual precipitation generally below 400 mm except at higher elevations. In contrast, Washington features strong orographic precipitation gradients, supporting some of the wettest forest ecosystems in the United States (https://prism.oregonstate.edu/). The study area encompasses major North American ecoregions, including the Great Basin and Western Cordillera cold deserts, the warm deserts of southern Nevada, and the marine West Coast forests of western

Washington. Vegetation types in Utah and Nevada include arid shrublands, sagebrush steppe, and higher-elevation pinyon-juniper woodlands and mixed conifer forests dominated by *Pinus flexilis*, *Abies concolor*, and *Pseudotsuga menziesii* (Miller et al., 2013). In contrast, Washington's temperate rainforests support some of the tallest and most biomass-rich tree species globally, shaped by historical low-intensity fire regimes and a century of active fire suppression (Keith et al., 2009; Palmer et al., 2019; Stine et al., 2014). This range in climate, vegetation structure, and ecosystem productivity provides a robust natural gradient for assessing the accuracy and generalizability of the *terraPulse* AGBD product across both low- and high-biomass environments.

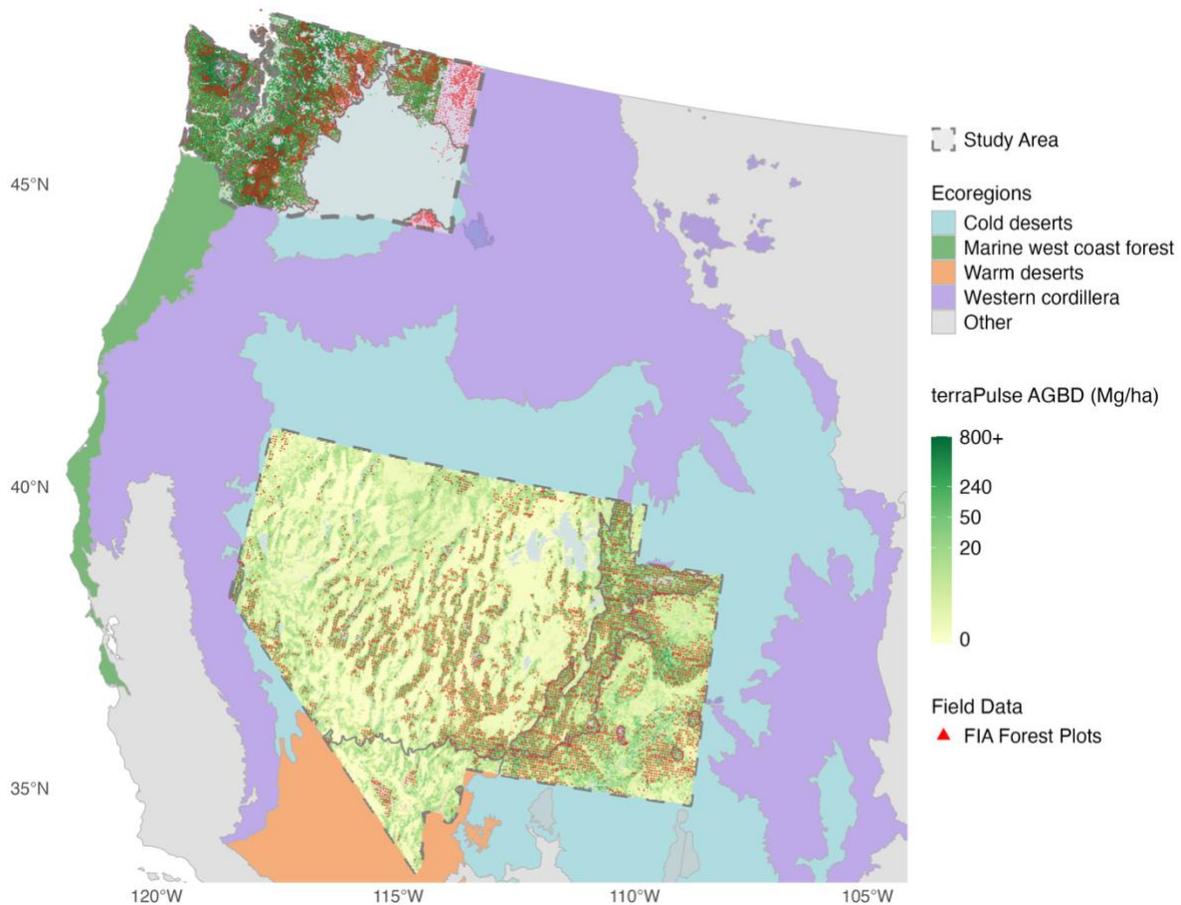

**Figure 1.** Overview of the study area and aboveground biomass density (AGBD) data. The background ecoregion polygons are based on the U.S. Environmental Protection Agency (EPA) Level 2 ecoregions (Dinerstein et al., 2017). The green gradient colors in the study area represent AGBD estimates for 2020

from terraPulse, Inc. The US Forest Service's Forest Inventory and Analysis (FIA) data (measure year ranging from 2000 to 2021) are used as reference AGBD data to validate the *terraPulse* estimates.

## 2.2 Remote-sensing based AGBD estimation

Annual, 30-meter resolution AGBD estimates within our study areas were produced by terraPulse, Inc. (Figure 1). The *terraPulse* dataset estimates AGBD in a spatio-temporal location $i$ (i.e., pixel and year) using a model $f$ based on remotely sensed variables $X$:

$$AGBD_i = f(X;\beta) + \varepsilon_i \qquad (1)$$

where $\beta$ is a set of empirically estimated parameters and $\varepsilon$ is the residual error. The function $f$ is fit by gradient-boosting regression trees (GBRTs; Ke et al., 2017), which are well-suited for handling large and heterogeneous datasets with complex, nonlinear relationships between response and covariates.

Training data were derived from GEDI footprint-level AGBD values (R. O. Dubayah et al., 2021) from 2019 to 2022. To resolve the limited spatial coverage of GEDI at high latitudes (51.6° N - 51.6° S), additional AGBD estimates derived from ICESAT-2 over the same period were incorporated into the training dataset. Model covariates $X$ included *terraPulse* tree cover (Sexton et al., 2013), forest stand age, deciduous-evergreen fractions, and annual mean and standard deviation Normalized Difference Vegetation Index (NDVI), as well as land cover type from the European Space Agency (ESA) Climate Change Initiative (CCI) and elevation, slope, and aspect derived from the ALOS World 3D-30m (AW3D30) dataset (https://www.eorc.jaxa.jp/ALOS/en/dataset/aw3d30/aw3d30_e.htm). Two models were fitted to accommodate unmeasurable stand age beyond 40 years (i.e., the length of the Landsat satellite mission): one model including stand age and an alternate model excluding age from the set of covariates $X$. This generated two pairs of AGBD predictions, each including both the AGBD estimate and its uncertainty (as RMSE), the weighted mean of which was taken as the final biomass prediction in each pixel.

The models were fit based on 1° × 1° tiles of training data. To generate spatially smooth predictions across image tiles and between years, reference data from 3 × 3-tile windows were randomly sampled to produce a model for the center tile and year of each window. To generate temporally smooth predictions, reference data were also sampled over the temporal extent of available reference data. Water bodies were masked from AGBD estimates using the *terraPulse* annual inundation frequency dataset (Feng et al., 2016); pixels with >50% inundation frequency were masked as null values from the respective years of the final AGBD product.

## 2.3 Ground-based AGBD reference estimates

We used data from the FIA for ground reference AGBD estimates (Figure 1). The FIA program has conducted systematic forest inventories across Utah, Nevada, and Washington since the early 2000s, with a sampling intensity of approximately one plot per 2,400 ha (McRoberts, 2006). FIA plots include species-level tree measurements—such as height, diameter, and live/dead status—from which total aboveground biomass is calculated using standard allometric equations (Burrill et al., 2021; Westfall et al., 2024). The FIA records the location of each plot, and each individual tree record can be linked to the respective plot. Therefore, we were able to calculate the AGBD of each FIA plot. The data are publicly accessible in SQLite database or separate CSV files from FIA DataMart (https://apps.fs.usda.gov/fia/datamart/datamart.html)

Because FIA plot coordinates are spatially perturbed to preserve privacy—via "fuzzing" within 0.8–1.5 km and random "swapping" of up to 20% of private plots within a county (Burrill et al., 2021)—direct pixel-to-plot comparisons with remote sensing data are infeasible. To mitigate this, we used two forms of spatially aggregated FIA estimates (Figure 2). First, we employed a hexagon-scale dataset developed by Menlove and Healey (Menlove & Healey, 2020), which provides FIA-derived AGBD estimates at the scale of Environmental Monitoring and Assessment Program (EMAP) hexagons (~64,000 ha each) (Menlove & Healey, 2020; White et al., 1992). These estimates are spatially continuous across the conterminous U.S. and incorporate all available FIA plots within each hexagon, including those on non-forest lands. Variables reported in the dataset include hexagon-scale mean AGBD, the number of contributing plots, sampling error, forest cover proportion, and mean inventory year. We used estimates based on the component ratio method (CRM) (Heath et al., 2009), focusing on the live-tree carbon pool to maintain consistency with the *terraPulse* AGBD product. While this scale represents the finest spatial resolution achievable with FIA's sampling density (typically ~25–30 plots per hexagon) (Menlove & Healey, 2020), the use of temporally pooled plots introduces uncertainty in capturing recent biomass dynamics, especially in regions with rapid disturbance or regrowth.

To address this limitation, we also produced county-level AGBD estimates using FIA-defined evaluation identifiers, which represent the subset of plots appropriate for producing estimates for specific inventory years (Bechtold & Patterson, 2015; Burrill et al., 2021). In Utah and Nevada, counties are official FIA estimation units. In Washington, where counties are not predefined as estimation units, we manually delineated counties as estimation units for consistency. Each county in Washington was stratified into forest and non-forest land cover classes—consistent with the approach FIA used for the other two states—using the most recent U.S. Geological Survey National Land Cover Database (NLCD; https://www.usgs.gov/centers/eros/science/national-land-cover-database), and AGBD estimates were produced within each stratum following standard FIA post-stratified estimation procedures. This county-scale approach provides temporally aligned reference values and complements the hexagon-scale data by enabling more precise comparisons in regions with recent forest change.

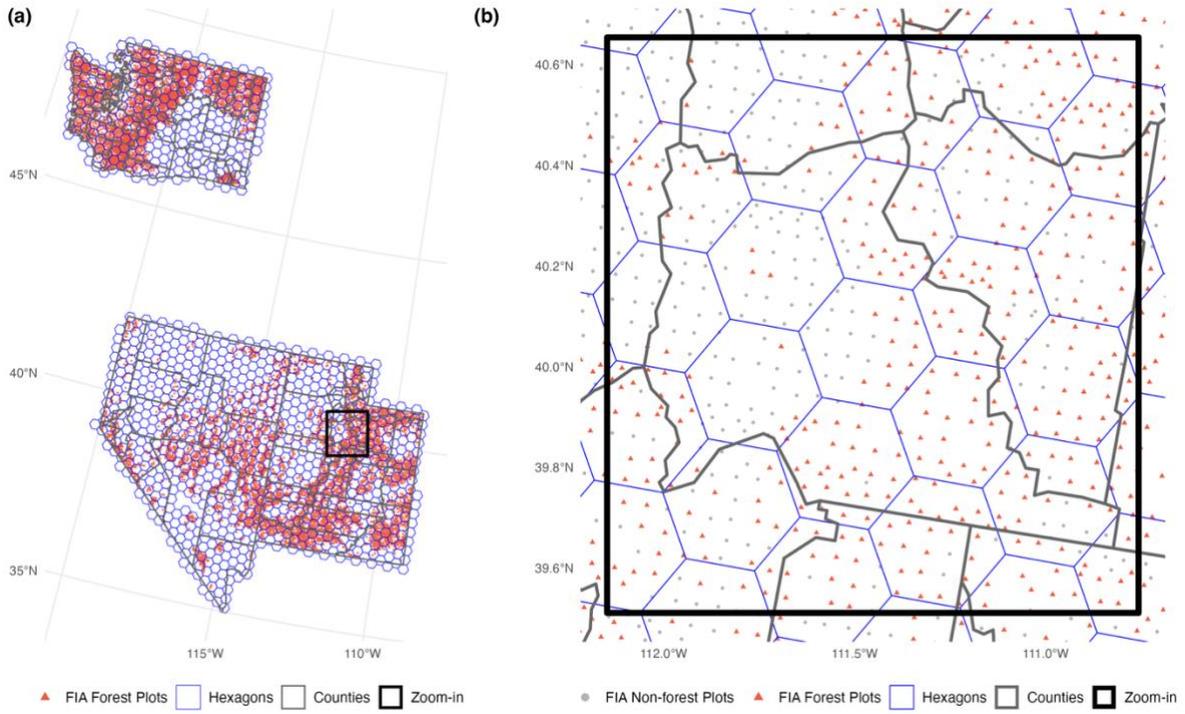

**Figure 2**. The FIA plots and two aggregation scales for validating the *terraPulse* biomass estimates in our study area. The two aggregation scales are 64,000-hectare hexagons (blue hexagons) and US counties (gray polygons). (a) The overview of the FIA forest plot distribution and the two aggregation scales. (b) The zoomed-in area is in the northern part of Utah.

## 2.4 Validation

To compare the *terraPulse* and FIA AGBD datasets, we aggregated the *terraPulse* pixel-level data to match the aggregated FIA estimates. The *terraPulse* AGBD pixel estimates were firstly reprojected to the Albers Equal Area Conic projection, and their values were then aggregated to hexagons and counties using zonal statistics. For the hexagon-scale comparison, we matched each state to the *terraPulse* AGBD map corresponding to the mean FIA inventory year of its hexagons. In Utah and Nevada, inventory years aligned within the *terraPulse* record (2001–2021). In Washington, however, the mean inventory year was 2014—outside the current *terraPulse* data range—so we used 2020 as a proxy for validation. Similarly for the county-scale comparison, we matched *terraPulse* and FIA estimates for all years between 2000 and 2021 where both were available, focusing primarily on the year 2020. Counties in Washington not fully covered by *terraPulse*—primarily those east of 118° W and in the central cold desert—were excluded from analysis (n = 14 counties removed).

To minimize bias from incomplete spatial coverage, we filtered out hexagons with insufficient valid *terraPulse* pixel data. For each state, we computed a pixel count threshold:

$$T_n = \bar{n} - \sigma_n \tag{2}$$

where $\bar{n}$ is the mean pixel count in the hexagons for the state, and $\sigma_n$ is the standard deviation of the pixel count in the hexagons. Hexagons with pixel counts below $T_n$ were excluded. After filtering, 959 of the original 1,190 hexagon units remained.

We evaluated the *terraPulse* AGBD estimates both statistically and spatially. To assess overall agreement, we compared the distributions and summary statistics of the *terraPulse* and FIA AGBD estimates. To visualize the distribution of the two datasets, we randomly sampled *terraPulse* values from each 10 × 10 pixel window to reduce data volume before producing histograms. We then assessed the spatial agreement of AGBD estimates from *terraPulse* and FIA using a t-test for each aggregation unit, following the equation:

$$t = \frac{\hat{\mu}_{terraPulse} - \hat{\mu}_{FIA}}{\sqrt{v(\hat{\mu}_{terraPulse}) + v(\hat{\mu}_{FIA})}} \tag{3}$$

where $\hat{\mu}_{FIA}$ and $v(\hat{\mu}_{FIA})$ are the estimated mean AGBD and variance of the sample mean from the FIA design-based plots within each estimation unit, and $\hat{\mu}_{terraPulse}$ and $v(\hat{\mu}_{terraPulse})$ are the corresponding values from the *terraPulse* AGBD estimates. Equation (3) is a test of the hypothesis whether the *terraPulse* predictions are equivalent to the FIA unbiased estimates. We did not report confidence intervals, as high variance can render them uninformative or misleading (R. Dubayah et al., 2022; Morey et al., 2016); instead, we interpreted t-statistics in relation to the critical values ±2: values within this range indicate the difference between *terraPulse* and FIA estimates is less likely to be significant. We also mapped the spatial distribution of t-statistics to assess regional patterns of agreement and discrepancy.

To evaluate model performance, we computed the coefficient of determination ($R^2$), root mean square error (RMSE), Pearson's correlation coefficient (r), and the slope and intercept of linear regressions comparing *terraPulse* and FIA values. We also calculated the index of agreement $d_r$ (Willmott et al., 2012), a bounded metric (−1 to 1) quantifying predictive concordance, defined as:

$$d_r = \begin{cases} 1 - \dfrac{\sum_{i=1}^{n}|P_i - O_i|}{c\sum_{i=1}^{n}|O_i - \bar{O}|}, \text{when} \\ \sum_{i=1}^{n}|P_i - O_i| \leq c\sum_{i=1}^{n}|O_i - \bar{O}| \\ \dfrac{c\sum_{i=1}^{n}|O_i - \bar{O}|}{\sum_{i=1}^{n}|P_i - O_i|} - 1, \text{when} \\ \sum_{i=1}^{n}|P_i - O_i| > c\sum_{i=1}^{n}|O_i - \bar{O}| \end{cases} \quad (4)$$

where $c=2$, $P_i$ and $O_i$ are paired model predicted and reference values, and $\bar{O}$ is the mean of the reference values. $d_r$ is normalized between -1 and 1, quantifying how closely the predicted values ($P_i$) match the reference values ($O_i$) relative to the natural variation in the reference data. A higher $d_r$ indicates stronger agreement. All data analyses were performed in R v.4.4.3 (R Core Team, 2025).

# 3 Results

## 3.1 Distribution of FIA and *terraPulse* AGBD estimates

The distributions of both FIA plot-scale and the original *terraPulse* AGBD estimates were strongly right-skewed, with a substantial portion of values concentrated at the lower end of the biomass range (Figure 3). The mean AGBD across FIA plots was 227 Mg ha$^{-1}$, with a median of 138 Mg ha$^{-1}$. The estimates ranged up to a maximum of 4,372.9 Mg ha$^{-1}$, with the 98th percentile occurring at 1,155.1 Mg ha$^{-1}$. More than 50% of FIA plot values were less than 100 Mg ha$^{-1}$, consistent with FIA's sampling focus on forested land and its inclusion of regenerating or sparsely stocked plots. The *terraPulse* pixel-level AGBD estimates were similarly distributed but more heavily weighted toward low-biomass values. Approximately 80% of *terraPulse* estimates fell below 50 Mg ha$^{-1}$, the mean and median were 48.4 Mg ha$^{-1}$ and 6.8 Mg ha$^{-1}$, respectively. The 98th percentile of the distribution was 361.7 Mg ha$^{-1}$, and the maximum value reached 1,110.3 Mg ha$^{-1}$—lower than the FIA maximum but still capturing 98% of the FIA AGBD range.

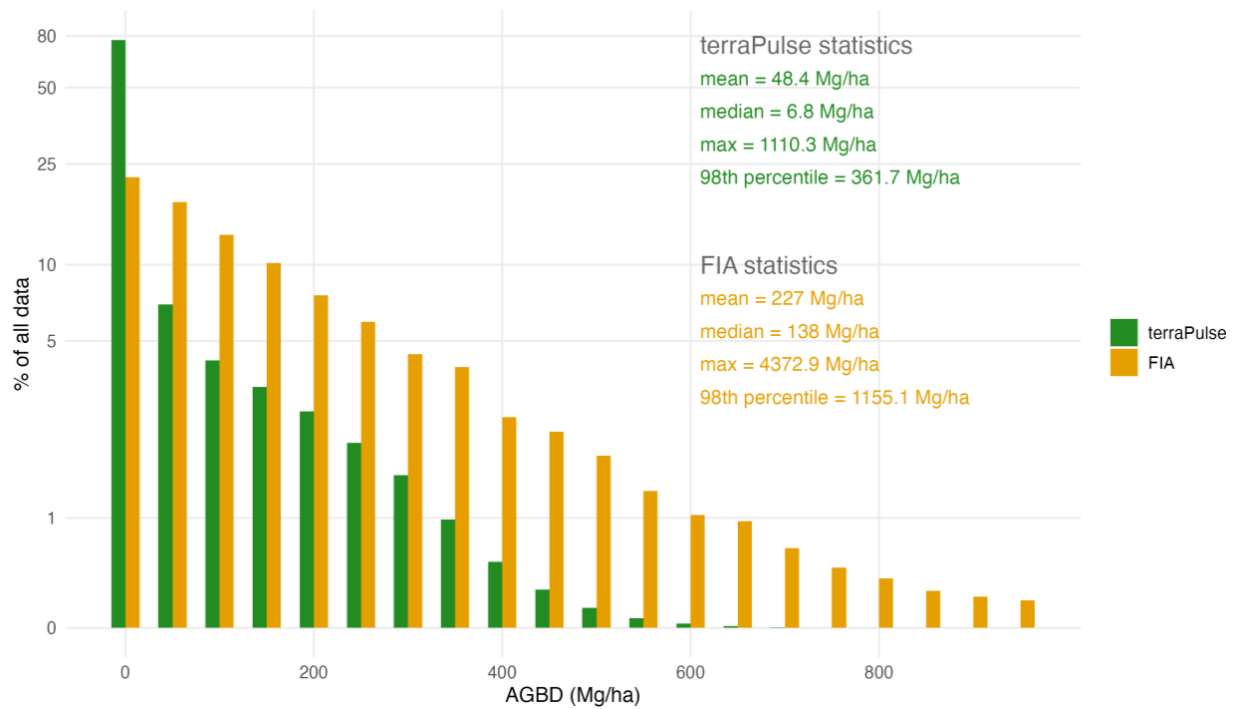

**Figure 3**. Distributions and summary statistics of the original *terraPulse* AGBD and FIA plot-scale AGBD estimates in our study area in 2020. The width of each bin is 50 Mg ha$^{-1}$. The y-axis was log-transformed to better visualize the tail of the right-skewed distribution.

These differences in distribution revealed the distinct domains of the two datasets: FIA measurements are restricted to forested areas, while *terraPulse* provided spatially continuous estimates, including in non-forested or low-biomass regions. The observed saturation in *terraPulse* values at the upper tail also likely reflected limits in the optical covariates used for model prediction, particularly in dense or structurally complex forest types.

## 3.2 Hexagon-scale AGBD comparison

We compared *terraPulse* and FIA AGBD estimates at the 64,000-hectare hexagons scale (Figure 4). The spatial distribution of AGBD from *terraPulse* closely mirrored that of FIA, with both products exhibiting similar regional biomass patterns (Figure 4a-c). However, the FIA map showed more hexagons with near-zero AGBD, particularly in areas dominated by non-forest land cover (Figure 4a,b). This resulted from FIA's design-based exclusion of biomass estimates in areas without forest plots, effectively assigning zero values in unsampled regions (see Figure 2). Yet spatial agreement between the two datasets was high, with t-statistics in most hexagons between –2 and 2, suggesting that differences between *terraPulse* and FIA AGBD were less likely to be statistically significant across the majority of the study area (Figure 4c). Discrepancies exhibited clear spatial structure as well: positive t-statistics tended to occur in low-AGBD regions where FIA estimates were near zero and *terraPulse* assigned small but nonzero biomass values, while negative t-statistics clustered in high-AGBD regions—particularly forested zones in western Washington—where *terraPulse* tended to underestimate compared to FIA.

Statistical agreement between coincident FIA and *terraPulse* AGBD values was likewise strong (Figure 4d,e). Most differences between the terraPulse and FIA AGBD were positive and clustered near zero (Figure 4d). This indicated that a part of terraPulse estimates tended to be larger than FIA estimates, but most of the differences were small. The linear regression of *terraPulse* on FIA reference AGBD (Figure 4e) yielded an $R^2$ of 0.88, with a slope of 0.99 (95% CI: 0.97–1.02, $p < 0.01$) and intercept of 9.79 Mg ha$^{-1}$ (95% CI: 8.05–11.53, $p < 0.01$). The slope near unity indicated a high degree of fidelity in *terraPulse* estimates across the AGBD range. The intercept reflected a modest tendency for *terraPulse* to assign small nonzero values where FIA estimates were zero, again likely due to coverage in sparsely vegetated areas. The RMSE of 26.68 Mg ha$^{-1}$, Pearson's r = 0.94, and index of agreement ($d_r$ = 0.82) reflected strong agreement between the datasets. Notably, the largest discrepancies between datasets occurred in hexagons corresponding to the marine West Coast forest ecoregion, where FIA plots reported very high AGBD and *terraPulse* predictions tended to saturate. The Q–Q plot (Figure 4e inset) further highlighted this gradient: *terraPulse* tended to overestimate FIA AGBD in low-biomass hexagons (<250 Mg ha$^{-1}$) and underestimated it in high-biomass hexagons (>300 Mg ha$^{-1}$).

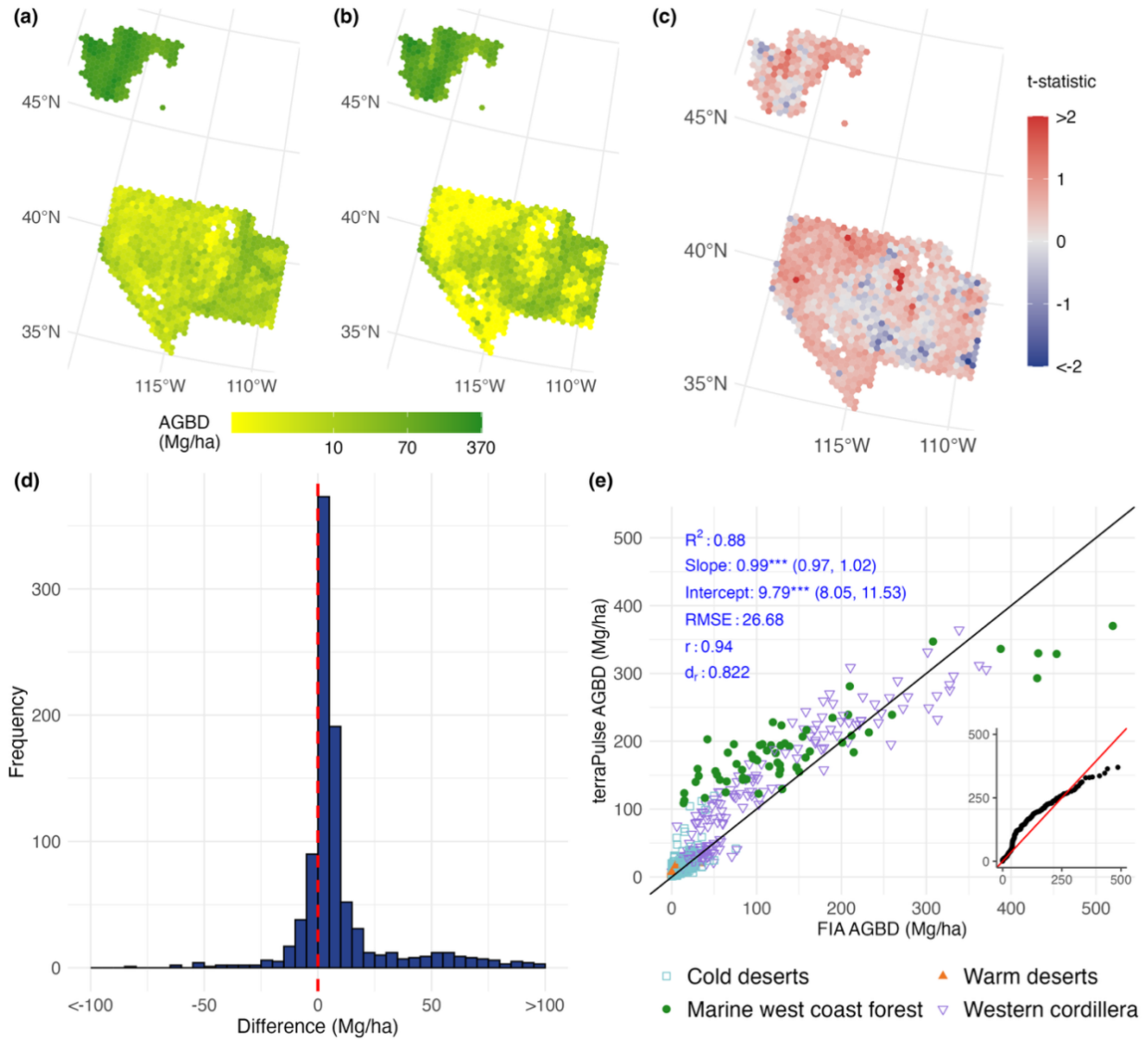

**Figure 4**. The 64,000-ha hexagon-scale comparison between *terraPulse* AGBD and FIA AGBD. (a) and (b) AGBD estimates of each hexagon from (a) *terraPulse* and (b) FIA. (c) The map of the t-statistic for the difference between the *terraPulse* and FIA AGBD estimates (*terraPulse* - FIA). Values roughly between -2 and 2 indicate that the differences are less likely to be significant. (d) Histogram of the differences between *terraPulse* and FIA estimates (*terraPulse* - FIA) for each hexagon. (e) Scatter plot of AGBD estimates from *terraPulse* and FIA for each hexagon. The inset shows the Q-Q plot of *terraPulse* vs. FIA AGBD, with axes aligned with the axes of the scatter plot. The solid line is a 1:1 line. $R^2$, slope, and intercept are calculated from linear regression between the two estimates. Asterisk counts indicate the significance of the slope and intercept, with more than two asterisks indicating $P < 0.01$. Values in the parentheses are 95% confidence intervals. RMSE is the root mean square error between the two estimates;

r is the Pearson correlation coefficient; $d_r$ is the index of agreement calculated following Willmott et al., 2012.

## 3.3 County-scale AGBD comparison

Agreement between *terraPulse* and FIA AGBD estimates increased at the county scale (Figure 5). Spatial distributions of AGBD from *terraPulse* (Figure 5a) and FIA (Figure 5b) closely matched, with consistent regional patterns in biomass density. In Utah and Nevada, FIA estimates were generally higher than those from *terraPulse*, particularly in moderately forested counties. However, in counties with very low AGBD—typically dominated by desert or shrubland—the two datasets converged, reflecting minimal biomass presence in both sources. The spatial distribution of t-statistics (Figure 5c) revealed that all counties fell within the nonsignificant range ($-2 < t < 2$), indicating that differences between *terraPulse* and FIA estimates were less likely to be statistically significant at this scale. Despite this, meaningful spatial patterns were evident. Counties with positive t-statistics tended to coincide with low-biomass areas, indicating that *terraPulse* estimates tended to be higher compared to FIA in these areas. Conversely, counties with negative t-statistics aligned with higher-biomass regions, suggesting that *terraPulse* slightly underestimated AGBD relative to FIA in those areas.

Residual analysis and Regression further confirmed the overall agreement (Figure 5d,e). Most differences between *terraPulse* and FIA AGBD were positive and clustered near zero (Figure 5d). This indicated that a part of the terraPulse estimates tended to be larger than the FIA estimates, but most of the differences were small. The linear regression (Figure 5e) yielded an $R^2$ of 0.90, with a slope of 1.07 (95% CI: 0.98–1.15, $p < 0.01$) and an intercept of 12.56 Mg ha$^{-1}$ (95% CI: 4.16–20.96, $p < 0.01$), indicating that *terraPulse* predictions tracked FIA estimates closely and tended to slightly overpredict at the low end of the biomass range. The RMSE was 32.62 Mg ha$^{-1}$, and r was 0.95, indicating strong linear association. The index of agreement ($d_r$ = 0.85) also reflected high consistency in spatial and quantitative AGBD estimation.

Most of the observed deviations occurred in Washington counties with mid-range AGBD values, where forest structure and cover variability were high. The Q–Q plot (Figure 5e inset) revealed that *terraPulse* tended to overestimate AGBD in counties with median biomass levels (50–200 Mg ha$^{-1}$), while performance at both the low and high ends of the biomass distribution was more closely aligned with FIA. These findings suggested that the *terraPulse* product generalized well across biomass regimes but may have exhibited moderate bias in structurally heterogeneous, mid-biomass regions.

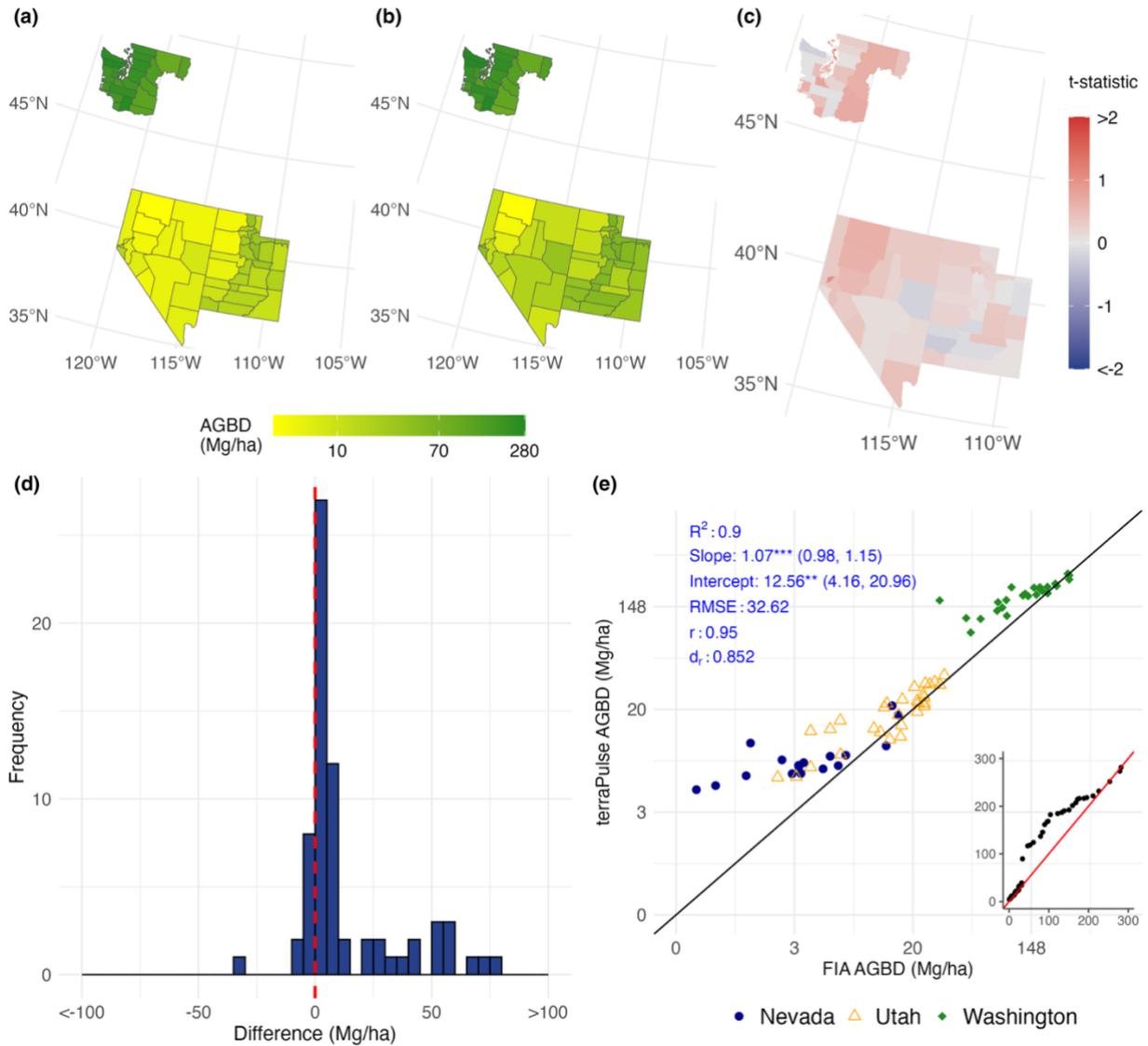

**Figure 5**. County-scale comparison between *terraPulse* AGBD and FIA AGBD for 2020. (a) and (b) AGBD estimates of each county from (a) *terraPulse* and (b) FIA. (c) Map of the t-statistic for the difference between *terraPulse* and FIA AGBD estimates. (d) Histogram of the differences between *terraPulse* and FIA estimates (*terraPulse* - FIA) for each county. (e) Scatter plot of AGBD estimates from *terraPulse* and FIA for each hexagon, with inset showing the Q-Q plot of *terraPulse* vs. FIA AGBD. The axes for the scatter plot were log-transformed to visualize the skewed distribution.

# 4 Discussion

In this study, we validated the *terraPulse* aboveground biomass density (AGBD) dataset using independent field estimates from the U.S. Forest Service's Forest Inventory and Analysis (FIA) program across three western U.S. states—Utah, Nevada, and Washington. To address the spatial uncertainty in FIA plot locations, we implemented a two-scale validation framework: 64,000-hectare hexagons and U.S. counties. These scales provided complementary perspectives, with hexagons enabling fine-resolution validation and counties allowing more temporally aligned comparisons using FIA-defined evaluation units.

Validation results demonstrated strong agreement between *terraPulse* and FIA AGBD estimates. At the hexagon scale, *terraPulse* showed high accuracy across most of the AGBD gradient, with slight overestimation at the lower end and underestimation at the high end, likely reflecting saturation in the optical remote sensing covariates. At the county scale, *terraPulse* exhibited strong overall accuracy, though discrepancies were more apparent in mid-range AGBD areas. Despite these differences, the validation confirmed that *terraPulse* provided spatially consistent, wall-to-wall AGBD estimates, and introduced a robust and replicable framework for evaluating remote sensing biomass products using FIA reference data.

## 4.1 Advantages of the *terraPulse* AGBD product

The *terraPulse* AGBD product differs from previous remote sensing biomass datasets in several important ways. First, it offers broad spatial and temporal coverage, providing up to annual global AGBD estimates from 1984 to the present. This is made possible by integrating GEDI and ICESat-2 LiDAR measurements with multiple ecological covariates, including tree cover, vegetation indices, and topography. Second, its 30-meter resolution enables finer-scale biomass monitoring than many existing products, while remaining ecologically interpretable by avoiding sub-canopy pixel sizes (Duncanson et al., 2025).

## 4.2 Interpretation of discrepancies

While *terraPulse* and FIA estimates aligned well overall, we observed discrepancies stemming from a combination of methodological, ecological, and spatial factors. The differences in distribution (Figure 3) reflected, in part, the broader spatial domain of *terraPulse* AGBD, which included low-biomass areas such as shrublands, grasslands, and deserts—regions excluded from FIA sampling due to the program's forest definition (≥10% tree canopy within the past 30 years) (Palmer et al., 2019). As a result, *terraPulse* included many more low-biomass pixels, while FIA assigned zero biomass in non-forest areas. This also explained *terraPulse*'s greater frequency of low AGBD values and its slight positive bias in low-biomass hexagons and counties.

In contrast, underestimation of AGBD by *terraPulse* in high-biomass regions—particularly in the marine West Coast forests—was likely due to saturation in optical indices and limits in LiDAR-derived training data. While *terraPulse* reached a maximum AGBD of ~1,100 Mg ha$^{-1}$, closely matching the 98th percentile of FIA estimates, it did not capture the highest FIA-reported values (~4,300 Mg ha$^{-1}$). This may reflect both model saturation and spatial averaging over 900 m² pixels, compared to FIA's smaller subplot-based estimates (~670 m²) (Burrill et al., 2021), which may be less subject to within-pixel averaging.

At the hexagon scale, *terraPulse* tended to overestimate AGBD in low-biomass regions and underestimate it in high-biomass areas. The overestimation may actually reflect improved characterization of non-forest biomass, while underestimation at the high end remains a known limitation in current remote sensing techniques. At the county scale, *terraPulse* performed better in high-AGBD counties, especially in Washington, likely due to spatial aggregation mitigating saturation effects and increasing signal stability through the inclusion of non-forest biomass.

Temporal mismatches also contributed to observed discrepancies, especially at the hexagon scale. For example, the average FIA inventory year for Washington was 2014, but was compared against the 2020 *terraPulse* dataset due to data availability, introducing potential biases from forest growth, harvest, or disturbance. At the county scale, temporal alignment was improved through the use of FIA evaluation identifiers. Additionally, FIA's differing treatment of estimation units across states—counties in Utah and Nevada versus user-defined units in Washington—may have introduced variability in the accuracy of aggregated reference estimates.

## 4.3 Limitations and future directions

Although this study demonstrates strong performance of the *terraPulse* AGBD product and the rigorous validation framework, several limitations remain. Most notably, the spatial resolution of the validation was limited to aggregated hexagon and county scales, owing to the location perturbation of FIA plots. As a result, the accuracy of *terraPulse* estimates at its native 30-meter resolution remains untested. Future validation efforts should incorporate plot-level datasets with precise geolocation and fine-grained sampling across the biomass gradient.

Another limitation involves inconsistencies in FIA biomass estimation methods. Beginning in late 2023, FIA transitioned from the Component Ratio Method (CRM) (Heath et al., 2009) to the National Scale Volume and Biomass Estimator (NSVB) (Westfall et al., 2024). Our county-scale validations used the updated NSVB estimates, while the hexagon-scale FIA dataset was based on CRM. Since CRM has been shown to underestimate biomass relative to NSVB (Chojnacky, 2012), this discrepancy may have influenced validation outcomes. Harmonizing allometric estimation methods across FIA products will be critical for future validation efforts.

# 5 Conclusions

The growing availability of satellite-derived biomass observations offers unprecedented opportunities for global AGBD monitoring. The *terraPulse* dataset demonstrated strong agreement with independent FIA reference data across a range of biomass regimes and ecological conditions. Its broad spatial coverage, long temporal record, and fine spatial resolution make it a valuable tool for carbon accounting, land management, and ecological assessment. This study introduced a rigorous, scalable, and replicable framework for validating remote-sensing AGBD products against FIA field data, accounting for spatial uncertainty and ecological variability. The successful validation of the *terraPulse* product underscores the potential for next-generation biomass monitoring systems to support national carbon inventories, ecosystem service assessments, and market-based climate mitigation programs.

# CRediT authorship contribution statement

**Xiuyu Cao**: Conceptualization, Methodology, Data curation, Formal analysis, Writing - original draft, Writing - review and editing. **Joseph O. Sexton:** Conceptualization, Methodology, Data curation, Writing - review and editing. **Panshi Wang:** Conceptualization, Methodology, Data curation, Writing - review and editing. **Dimitrios Gounaridis:** Supervision, Writing - review and editing. **Neil H. Carter:** Data curation, Writing - review and editing. **Kai Zhu:** Supervision, Conceptualization, Resources, Writing - review and editing.

# Acknowledgements

Research funding was generously provided by the NASA Biodiversity Program (grant no. 80NSSC21K1940), NSF (grant no. 2306198), USDA McIntire-Stennis Cooperative Forestry Research Program, and the University of Michigan Biosciences Initiative. The authors would like to thank members of the Zhu Lab at the University of Michigan for constructive comments on the manuscript.

# Data availability

We retrieved the US Forest Inventory and Analysis data from FIA DataMart (https://apps.fs.usda.gov/fia/datamart/datamart.html) in SQLite form on Feb 3, 2025. The US states and counties boundary data were retrieved using R package *maps*. The *terraPulse* aboveground biomass data are freely available in the NomadIQ mobile application (https://web.nomadiq.earth/layer-details/19). The R code and intermediate data are currently available on Figshare (https://figshare.com/s/37ace16e21037bf971b6) for peer review and will be permanently archived and made publicly available upon acceptance for publication.